\documentclass[a4paper,12pt]{article}

\usepackage{cite}

  \usepackage[pdftex]{graphicx}
\usepackage{xcolor}

\usepackage[cmex10]{amsmath}
\usepackage{amsfonts}

\usepackage{algorithmic}
\usepackage{algorithm}

\usepackage{array}

\usepackage[tight,normalsize,sf,SF]{subfigure}

\newcommand{\vc}[1]{\ensuremath{\boldsymbol{#1}}}
\newcommand{\mt}[1]{\ensuremath{\boldsymbol{#1}}}

\graphicspath{./figures/}

\begin{document}

\title{Sequential Monte Carlo Methods for State and Parameter
Estimation in Abruptly Changing Environments}

\author{Christopher Nemeth, Paul Fearnhead, Lyudmila Mihaylova}

\maketitle

\begin{abstract}
This paper develops a novel sequential Monte Carlo (SMC) approach for joint state and parameter estimation that can deal efficiently with abruptly changing parameters which is a common case when tracking maneuvering targets. The approach combines Bayesian methods for dealing with changepoints with methods for estimating static parameters within the SMC framework. The result is an approach which adaptively estimates the model parameters in accordance with changes to the target's trajectory. The developed approach is compared against the Interacting Multiple Model (IMM) filter for tracking a maneuvering target over a complex maneuvering scenario with nonlinear observations. In the IMM filter a large combination of models is required to account for unknown parameters. In contrast, the proposed approach circumvents the combinatorial complexity of applying multiple models in the IMM filter through Bayesian parameter estimation techniques. The developed approach is validated over complex maneuvering scenarios where both the system parameters and measurement noise parameters are unknown. Accurate estimation results are presented.
\end{abstract}

Sequential Monte Carlo methods, joint state and parameter
estimation, nonlinear systems, particle learning, tracking maneuvering
targets.

%

\section{Introduction}
%
%
%
%
State and parameter estimation for nonlinear systems is a challenging problem which arises in many practical areas, such as target tracking, control and communication systems, biological systems and many others. The main methods for state and parameter estimation or for parameter estimation only can be classified into two broad groups~\cite{Kantas2009,Doucet2009}:
Bayesian and Maximum Likelihood (ML) methods. Such methods may also be categorized as \emph{online} or \emph{offline} depending on whether the data are processed sequentially as new observations become available,
or processed in batches of observations. In ML estimation the optimal solution reduces to finding the estimate which maximizes the marginal likelihood of the observed data. The Bayesian approach, however, considers the parameters as random variables which are updated recursively using prior knowledge of the parameters (if available) and the measurement likelihood function. The approach proposed in this paper is an on-line Bayesian approach which uses sequential Monte Carlo (SMC) techniques.

Early attempts to solve the problem of estimating the parameters online involved selecting a prior distribution for the parameters and augmenting the state vector to include the unknown parameters. The parameters can then be estimated using the same filtering technique that is applied to the state. However, through successive time steps this approach quickly leads to particle degeneracy of the parameter space. The fixed nature of the parameters means that the particles which are sampled from the initial prior distribution do not vary with time, thus the same set of particles will be resampled with replacement from one time step to the next, reducing the number of unique particles, eventually resulting in multiple copies of the same particle. This creates a point mass approximation of the marginal posterior parameter distribution.

One solution to this problem is to perturb particles by adding artificial noise \cite{Gordon1993}. However, naively adding noise at each iteration can lead to overly diffuse distributions for the parameters, relative to the true posterior distribution \cite{West}. An improved and related approach is the Liu and West filter~\cite{West}. This filter uses \emph{kernel density estimation} to estimate the posterior distribution of the parameters, and in particular the idea of shrinkage to avoid
producing overly-diffuse approximations. An alternative approach to combat particle degeneracy is to use MCMC moves to sample new parameter values at each iteration. For some models this can be implemented efficiently, in an on-line setting, through the use of sufficient statistics \cite{Fearnhead2002,Storvik2002}. This class of methods has been recently termed \emph{particle learning}~\cite{Carvalho2009}.

Whilst these methods can work well with static parameters, the case with dynamically changing parameters remains still unresolved. Therefore, we are considering applications with time-varying parameters, and especially the cases where the parameter values can change abruptly at a small set of time-points \cite{Whiteley2011a}. A motivating application is in target tracking, where a maneuvering target typically has ``periods/segments'' of high and low maneuverability. The parameters, such as the turn-rate of a model for the target's dynamics will be constant within a segment but different between segments. This can be modeled through a time-varying parameter, but under the constraint that the parameter values are piecewise constant functions of time. We shall refer to this scenario as models with \emph{time-varying parameters} in the sequel.

Previous approaches to this problem include the jump Markov linear (JML) filter \cite{doucetetal:2001:pffseojmls}, where the parameters evolve according to a finite state Markov chain and the Interacting Multiple Model filter \cite{blom}.

In the IMM filter, numerous models are used (e.g. models for constant velocity and coordinated turn), each of which permit different fixed parameters, allowing the filter to switch between models depending on the motion of the target. The IMM filter has proven to be very successful for tracking highly maneuverable targets. However, the reliability of the IMM filter is dependent on the number and choice of models.

The IMM filter applies several proposed models (e.g. models for constant velocity and coordinated turn), each of which permit different fixed parameters, allowing the filter to account for various possible target behaviors. The IMM filter then merges the estimates of the various models based on their respective likelihood values to produce a single estimate of the target's state. This filter has proven to  be very successful for tracking highly maneuverable targets. However, the reliability of both the JML filter and the IMM filter are dependent on the \textit{a priori} tuning of the filters as neither of these filters aim to estimate the unknown parameters online. They also suffer a curse-of-dimensionality, if we wish to account for multiple unknown parameters, then the number of models required increases exponentially with the number of parameters.

The proposed approach accounts for time-varying parameters using changepoints, and then combining SMC approaches for changepoint models \cite{Fearnhead2007a}, \cite{yildirimandsinghanddoucet:2012:onlineem} with the standard SMC approaches for estimating static parameters ~\cite{Carvalho2009,West}. We call the resulting approach \emph{adaptive parameter estimation}. It allows learning of parameters within segments between changepoints, and also allows the parameter estimates to adapt and learn new values once a changepoint has occurred. Preliminary results were reported in~\cite{nemethandfearnheadandmihaylovaandvorley:2012:fusion} and \cite{nemethandfearnheadandmihaylovaandvorley:2012:ietconf}. This paper refines further the adaptive parameter estimation filters described in~\cite{nemethandfearnheadandmihaylovaandvorley:2012:fusion,nemethandfearnheadandmihaylovaandvorley:2012:ietconf} and presents a comparison with the IMM algorithm for complex maneuvering target scenarios.

The rest of the paper is organized as follows. Section \ref{sec:bayesian-filtering} presents the Bayesian formulation of the joint state and parameter estimation problem. Section \ref{sec:bayes-param-estim} describes
Bayesian approaches for joint state and parameter estimation. Section~\ref{sec:adapt-param-estim} presents the novel adaptive estimation algorithm. Section~\ref{sec:results} evaluates the performance of the developed
approach over two challenging scenarios with a maneuvering target. Finally, Section~\ref{sec:conclusion} generalizes the results and discusses future work.

\section{Bayesian Filtering}
\label{sec:bayesian-filtering}

A state-space model can be defined by two stochastic processes $\mt{X}_t$ and $\mt{Y}_t$. The process $\mt{X}_t$ is referred to as a \emph{hidden} or latent Markov process representing the state of interest at discrete time $t$, which takes values on the measurable space $\mathcal{X} \subseteq \mathbb{R}^{n_x}$. The stochastic process $\mt{Y}_t$ represents the observation process which takes values on the observation space $\mathcal{Y} \subseteq \mathbb{R}^{n_y}$, where observations are assumed to be dependent only on the current state $\mt{X}_t$ and independent of previous states $\mt{X}_{1:t-1}$, where $\mt{X}_{1:t-1}=\{ \mt{X}_{1},\mt{X}_{2},\ldots, \mt{X}_{t-1}\}$. We also assume that these stochastic processes are conditional upon the parameter vector $\boldsymbol{\theta}$, and that there exists a prior distribution, $p(\boldsymbol{\theta})$, for the parameter vector. The general state-space model is characterized by the densities:
\begin{eqnarray}
   \label{eq:1}
    \vc{X}_{t}|\{ \vc{x}_{0:t-1}, \vc{y}_{1:t-1}\}   & \sim  &  p( \vc{x}_t| \vc{x}_{t-1}, \boldsymbol{\theta}), \\
    \vc{Y}_t|\{\vc{x}_{0:t},\vc{y}_{1:t-1}\}  & \sim &  p(\vc{y}_t|\vc{x}_t, \boldsymbol{\theta}),
\end{eqnarray}
where the state model is conditional only on the previous state and the observations $\vc{y}_t$ are independent of previous observations conditional only on the state $\vc{x}_t$ at time $t$. Here $\vc{y}_{1:t-1}$ denotes the
measurements from time 1 to time $t-1$. 

In filtering, the aim is to estimate the hidden state at time point $t$ given a sequence of observations. This process requires the evaluation of the posterior probability density function
$p(\vc{x}_t,\boldsymbol{\theta}|\vc{y}_{1:t})$ of the hidden state vector and parameter vector conditional on the observations. Using Bayesian estimation techniques it is possible to evaluate the posterior density
recursively by first predicting the next state
\begin{equation*}
  \label{eq:5}
  p(\vc{x}_t, \boldsymbol{\theta}|\vc{y}_{1:t-1}) = \int p(\vc{x}_t|\vc{x}_{t-1},\boldsymbol{\theta})p(\vc{x}_{t-1},\boldsymbol{\theta}|\vc{y}_{1:t-1})d \vc{x}_{t-1}
\end{equation*}
and then updating this prediction to account for the most recent observation $\vc{y}_t$,
 \begin{equation}
  \label{eq:2}
   p(\vc{x}_t, \boldsymbol{\theta}|\vc{y}_{1:t}) = \frac{p(\vc{y}_t|\vc{x}_t, \boldsymbol{\theta})p(\vc{x}_t, \boldsymbol{\theta}|\vc{y}_{1:t-1})}{p(\vc{y}_t|\vc{y}_{1:t-1},\boldsymbol{\theta})},
 \end{equation}
  where
 \begin{equation}
  \label{eq:3}
   p(\vc{y}_t|\vc{y}_{1:t-1},\boldsymbol{\theta})=\int p(\vc{y}_t|\vc{x}_t,\boldsymbol{\theta})p(\vc{x}_t,\boldsymbol{\theta}|\vc{y}_{1:t-1})d \vc{x}_t.
 \end{equation}
is the normalizing constant. See \cite{Arulampalam2002} for a full details of this derivation.

Determining an analytic solution for the posterior distribution (\ref{eq:2}) is generally not possible due to the normalizing constant (\ref{eq:3}) being intractable. One exception is when the state-space is finite or
linear-Gaussian in which case an analytic solution can be found using a Kalman filter \cite{Kalman1960}. Generally, it is necessary to create an approximation of the posterior distribution, one such approach is through
sequential Monte Carlo methods, also known as particle filters.

Particle filters present a method for approximating a distribution using a discrete set of $N$ samples/particles with corresponding weights $\{\vc{x}_t^{(i)},\boldsymbol{\theta}^{(i)},w_t^{(i)}\}_{i=1}^{N}$ which create a
random measure characterizing the posterior distribution $p(\vc{x}_t,\boldsymbol{\theta}|\vc{y}_{1:t})$. The empirical distribution given by the particles and weights can then be used to approximate \eqref{eq:2} as
 \begin{equation}
  \label{eq:4}
   p(\vc{x}_t,\boldsymbol{\theta}|\vc{y}_{1:t}) \approx \sum_{i=1}^{N}w_t^{(i)}\delta((\vc{x}_t,\boldsymbol{\theta})-(\vc{x}_{t}^{(i)},\boldsymbol{\theta}^{(i)})),
 \end{equation}
 where $\delta(\cdot)$ is the Dirac delta function and each pair of particles $\vc{x}_t^{(i)}$ and $\boldsymbol{\theta}^{(i)}$ is given a weight $w_t^{(i)}$.

Using the empirical posterior distribution \eqref{eq:4} as an approximation to the true posterior distribution $p(\vc{x}_t,\boldsymbol{\theta}|\vc{y}_{1:t})$ it is possible to recursively update the posterior probability
density by \emph{propagating} and \emph{updating} the set of particles. The particles are propagated according to the dynamics of the system to create a predictive distribution of the hidden state at the next time step.
These particles are then updated by weighting each
particle based on the newest observations using principles from importance sampling \cite{Arulampalam2002}. Particle filtered approximations display inherent particle degeneracy throughout time due to an increase in the
variance of the importance weights \cite{Kong1994}. A popular solution to this problem is to discard particles with low (normalized) weights and duplicate particles with high (normalized) weights by using a resampling
technique \cite{Gordon1993}. Resampling the particles introduces Monte Carlo variation which produces poorer state estimation in the short term, but preserving particles with higher importance weights will provide greater
stability for the filter and produces better future estimates. There are several approaches to resampling particles, the simplest being simple multinomial resampling. However, improved resampling strategies such as
stratified resampling \cite{Carpenter1999} can minimize the introduced Monte Carlo variation (see \cite{Douc2005} for a review of resampling
strategies). In this paper we will use the systematic resampling technique \cite{Kitagawa1996a} which minimizes Monte Carlo variation and runs in $\mathcal{O}(N)$ time.

The next section describes important Bayesian approaches for state and parameter estimation: the auxiliary particle filter (APF) ~\cite{Pitt1999a} and particle learning techniques~\cite{Carvalho2009,Storvik2002,West} which
we use as a starting point to develop a novel adaptive Bayesian approach for state and parameter estimation.

\section{Bayesian State and Parameter Estimation}
\label{sec:bayes-param-estim}

\subsection{Auxiliary Particle Filter}
\label{sec:auxil-part-filt}
The original particle filter proposed by \cite{Gordon1993} suggests that the state particles $\{\vc{x}_t^{(i)}\}_{i=1}^{N}$ should be sampled from
the transition density $p(\vc{x}_t|\vc{x}_{t-1},\boldsymbol{\theta})$ and then weighted against the newest observation, which we shall refer to as
 \textit{propagate} - \textit{resample}. However, following this approach can lead to poor state estimates as the particles which are sampled from the transition density do not take account of the newest observations
 $\vc{y}_t$.
Ideally the state particles $\vc{x}_t^{(i)}$ would be sampled from the optimal importance distribution $p(\vc{x}_t|\vc{x}_{t-1},\vc{y}_t,\boldsymbol{\theta})$, which can be proven to be optimal \cite{Doucet2000} in
the sense that when applied it will minimize the variance of the importance weights. Sampling from the optimal importance distribution is generally not possible due to reasons of intractability.
The auxiliary filter as proposed by Pitt and Shephard~\cite{Pitt1999a}, offers an intuitive solution to this problem by resampling particles based on their predictive likelihood $p(\vc{y}_t|\vc{x}_{t-1},\boldsymbol{\theta})$, thus accounting for the newest observations $\vc{y}_t$ before the particles are propagated. This method can be viewed as a \textit{resample} - \textit{propagate} filter.

This filter can be considered as a general filter from which simpler particle filters are derived as special cases. Consider a modified posterior density $p(\vc{x}_t,\boldsymbol{\theta},k|\vc{y}_{1:t})$ of both state $\vc{x}_t$, parameter $\boldsymbol{\theta}$ and auxiliary variables $k$, where $k$ is the index of the particle at $t-1$. Applying Bayes theorem it can be shown that up to proportionality the target distribution is given by
\begin{equation}
\label{eq:11}
  p(\vc{x}_t,\boldsymbol{\theta},k|\vc{y}_{1:t}) \propto p(\vc{y}_t|\vc{x}_t,\boldsymbol{\theta}^{(k)})p(\vc{x}_t|\vc{x}_{t-1}^{(k)},\boldsymbol{\theta}^{(k)})w_{t-1}^{(k)},
\end{equation}
however, $p(\vc{y}_t|\vc{x}_t,\boldsymbol{\theta}^{(k)})$ is unavailable so instead we can sample from the proposal distribution
\begin{equation*}
  q(\vc{x}_t,\boldsymbol{\theta},k|\vc{y}_{1:t}) \propto p(\vc{y}_t|g(\vc{x}_{t-1}^{(k)}),\boldsymbol{\theta}^{(k)})p(\vc{x}_t|\vc{x}_{t-1}^{(k)},\boldsymbol{\theta}^{(k)})w_{t-1}^{(k)}
\end{equation*}
where $g(\vc{x}_{t-1}^{(k)})$ characterizes $\vc{x}_t$ given $\vc{x}_{t-1}^{(k)}$, usually we choose $g(\vc{x}_{t-1}^{(k)})=\mathbb{E}[\vc{X}_t|\vc{x}_{t-1}^{(k)},\boldsymbol{\theta}^{(k)}]$. Estimates of the posterior density $p(\vc{x}_t,\boldsymbol{\theta}|\vc{y}_{1:t})$ are given from the marginalized form of the density $p(\vc{x}_t,\boldsymbol{\theta},k|\vc{y}_{1:t})$ by omitting the auxiliary variable. Finally the importance sampling
weights which are given by the ratio of the target and proposal distributions, simplify to
\begin{equation*}
  w_t \propto \frac{p(\vc{y}_t|\vc{x}_{t},\boldsymbol{\theta}^{(k)})}{p(\vc{y}_t|g(\vc{x}_{t-1}^{(k)},\boldsymbol{\theta}^{(k)}))}.
\end{equation*}

\subsection{Particle Learning}
\label{sec:particle-learning}

Gilks and Berzuini~\cite{Gilks2001} proposed a Bayesian approach to parameter estimation based on Markov chain Monte Carlo (MCMC) steps, where the entire history of the states and the observations is used to update
the vector of unknown parameters $p(\boldsymbol{\theta}| \vc{x}_{0:t},\vc{y}_{1:t})$. The complexity of this approach grows in time and it suffers from the curse of dimensionality \cite{Bengtsson2008}.

Sampling parameters from the posterior parameter distribution $p(\boldsymbol{\theta}|\vc{x}_{0:t},\vc{y}_{1:t})$ becomes computationally more difficult as the time $t$ increases.
For some models a solution to this problem is to summarize the history of the states $\vc{x}_{0:t}$ and observations $\vc{y}_{1:t}$ via a set of low-dimensional sufficient statistic $\vc{s}_t$
\cite{Fearnhead2002,Storvik2002}.
 We define $\vc{s}_t$ to be \emph{sufficient statistic} if all the information from the states and observations can be determined through it,
 (i.e. $p(\boldsymbol{\theta}|\vc{x}_{0:t},\vc{y}_{1:t})=p(\boldsymbol{\theta}|s_t)$).
The sufficient statistic should be chosen such that it can be updated recursively as new states and observations become available $\vc{s}_t=\mathcal{S}_t(\vc{s}_{t-1},\vc{x}_t,\vc{y}_t)$.
It is possible to determine whether a function $\vc{s}_t$ is sufficient by the factorization theorem \cite{lindgren93}, which states that a function $\vc{s}_t$ is sufficient if there exist
functions $k_1(\cdot)$ and $k_2(\cdot)$ such that
\begin{equation}
  \label{eq:10}
  p(\boldsymbol{\theta}, \vc{x}_{0:t},\vc{y}_{1:t}) = k_1(\boldsymbol{\theta}_t,\mathcal{S}_t(\vc{s}_{t-1},\vc{x}_{t},\vc{y}_{t}))k_2(\vc{x}_{0:t},\vc{y}_{1:t}).
\end{equation}

The particle learning filter of Carvalho et al.~\cite{Carvalho2009} can be viewed as an extension to the works of Fearnhead \cite{Fearnhead2002} and Storvik \cite{Storvik2002} where sufficient statistics are used to
recursively update the posterior parameter distribution.
Particle learning differs from previous sufficient statistic approaches in that it is based on the auxiliary particle filter which works within the \textit{resample} - \textit{propagate} framework.
This approach produces better proposal distributions which more closely approximate the optimal proposal distribution, thus producing better state and parameter estimates. Particle learning also creates sufficient
statistics for the states when possible. This reduces the variance of the sample weights and is often referred to as Rao-Blackwellization.\\

\noindent The particle learning filter~\cite{Carvalho2009} is summarized in Algorithm~\ref{alg1}.
\begin{algorithm}
\caption{Particle Learning Filter}          
\label{alg1}                           
\begin{algorithmic}                    
    \STATE Sample particles $\{\vc{x}_{t-1}^{(i)},\boldsymbol{\theta}^{(i)}\}_{i=1}^{N}$ \\
    \quad \ \ with weights $w_{t}^{(i)} \propto p(\vc{y}_t|\boldsymbol{\mu}_{t}^{(i)},\boldsymbol{\theta}^{(i)})$ \\
    \quad \ \ where $\boldsymbol{\mu}_t^{(i)}=\mathbb{E}[\vc{x}_{t}|\vc{x}_{t-1}^{(i)},\boldsymbol{\theta}^{(i)}]$.
    \FOR {$i=1,\ldots, N$}
  \STATE Propagate state particles $\vc{x}_t^{(i)} \sim p(\vc{x}_t|\vc{x}_{t-1}^{(i)},\boldsymbol{\theta}^{(i)})$
    \STATE Update sufficient statistics with the newest \\
    \quad \ \ state and observation\\
    \quad\quad\quad\quad\quad\quad $\vc{s}_t^{(i)}=\mathcal{S}_t(\vc{s}_{t-1}^{(i)},\vc{x}_{t}^{(i)},\vc{y}_{t})$
    \STATE Sample new parameter values  \\
    \quad\quad\quad\quad\quad\quad $\boldsymbol{\theta}^{(i)} \sim p(\boldsymbol{\theta}|\vc{s}_t^{(i)})$
    \ENDFOR
\end{algorithmic}
\end{algorithm}

\subsection{Liu and West Filter}
\label{sec:liu-west-filter}

The implementation of the particle learning filter is dependent on producing a closed form conjugate prior for the parameters in order to define a sufficient statistic structure. For many complex models finding a
closed form conjugate prior is not possible, therefore, it is necessary to approximate the posterior marginal parameter distribution in an alternative way. Liu and West propose~\cite{West} an approach for approximating
the posterior marginal parameter distribution through a kernel density approximation, where the marginal posterior parameter distribution is approximated as a mixture of multivariate Gaussian distributions.

Using Bayes theorem it is possible to determine the joint posterior distribution for the state and parameter
$p(\vc{x}_t,\boldsymbol{\theta}_t|\vc{y}_{1:t})$ as
\begin{align*}
     p(\vc{x}_t,\boldsymbol{\theta}|\vc{y}_{1:t}) &\propto p(\vc{y}_t|\vc{x}_t,\boldsymbol{\theta})p(\vc{x}_t,\boldsymbol{\theta}|\vc{y}_{1:t-1}) \\
     & \propto p(\vc{y}_t|\vc{x}_t,\boldsymbol{\theta})p(\vc{x}_t|\vc{y}_{1:t-1},\boldsymbol{\theta})p(\boldsymbol{\theta}|\vc{y}_{1:t-1}),
\end{align*}
where the parameters are explicitly dependent on the observations.

The Liu and West filter can be interpreted as a modification of the artificial noise approach of Gordon et al. \cite{Gordon1993} without the loss of information. The marginal posterior of the parameter distribution is
represented as a mixture
\begin{equation*}
  \label{eq:6}
  p(\boldsymbol{\theta}|\vc{y}_{1:t-1}) \approx \sum_{i=1}^{N}w_{t-1}^{(i)}\mathcal{N}(\boldsymbol{\theta}| \vc{m}_{t-1}^{(i)},h^2 \mt{V}_{t-1}),
\end{equation*}
where $\mathcal{N}(\boldsymbol{\theta}|\vc{m}_{t-1}^{(i)},h^2\mt{V}_{t-1})$ is a multivariate normal density with mean and variance, 
\begin{eqnarray}
  \label{eq:7}
\vc{m}_{t-1}^{(i)} &=& a \boldsymbol{\theta}^{(i)}+(1-a)\overline{\boldsymbol{\theta}},  \\
\label{eq:9} \mt{V}_{t-1} &=& \sum_{i=1}^{N}w_{t-1}^{(i)}(\boldsymbol{\theta}^{(i)}-\overline{\boldsymbol{\theta}})(\boldsymbol{\theta}^{(i)}-\overline{\boldsymbol{\theta}})^\top,
\end{eqnarray}
where
$\overline{\boldsymbol{\theta}}=\sum_{i=1}^{N}w_{t-1}^{(i)}\boldsymbol{\theta}^{(i)}$ and $\mt{V}_{t-1}$ are the Monte Carlo posterior mean and variance of $\boldsymbol{\theta}$, respectively. The kernel smoothing parameter is denoted $h^2$ with shrinkage parameter $a = \sqrt{1-h^2}$ (discussed below) and $^\top$ as the transpose operation.

Standard kernel smoothing approximations suggest that kernel components should be centered around the parameter estimates, $\vc{m}_{t-1}^{(i)}=\boldsymbol{\theta}^{(i)}$. However, this approach can lead to overly-dispersed posterior distributions as the variance of the overall mixture is $(1+h^2) \mt{V}_{t-1}$ and therefore larger than the true variance $\mt{V}_{t-1}$.
The overly dispersed approximation for the posterior $p(\boldsymbol{\theta}|\vc{y}_{1:t-1})$ at time $t-1$ will lead to an overly-dispersed posterior $p(\boldsymbol{\theta}|\vc{y}_{1:t})$ at time $t$, which will grow with time. West~\cite{West1993} proposed a shrinkage step to correct for the over-dispersion by taking the kernel locations as in \eqref{eq:7}, where the shrinkage parameter $a$ corrects for the over-dispersion by pushing particles $\boldsymbol{\theta}^{(i)}$ back towards their overall mean. This results in a multivariate mixture distribution which retains $\overline{\boldsymbol{\theta}}$ as the overall mean with correct variance $\vc{V}_{t-1}$.

The Liu and West filter~\cite{West} is summarized in Algorithm \ref{alg2}.
\begin{algorithm}
\caption{Liu and West Filter}
\label{alg2}
  \begin{algorithmic}
   \STATE  Sample particles $\{ \vc{x}_{t-1}^{(i)}, \boldsymbol{\theta}^{(i)}\}_{i=1}^{N}$ \\
    \quad \ \ with weights $w_{t} \propto w^{(i)}_{t-1}p(\vc{y}_t|\boldsymbol{\mu}_{t}^{(i)},\vc{m}_{t-1}^{(i)})$ \\
    \quad \ \ where $\boldsymbol{\mu}_t=\mathbb{E}[\vc{x}_{t}|\vc{x}_{t-1}^{(i)},\boldsymbol{\theta}^{(i)}]$ and $\vc{m}_{t-1}^{(i)}$ \\
     \quad \ \ is given in \eqref{eq:7}
    \FOR {$ i=1,\ldots, N$}
    \STATE  Parameters are sampled from the kernel density \\
    \quad\quad $\boldsymbol{\theta}^{(i)} \sim \mathcal{N}(\boldsymbol{\theta}| \vc{m}_{t-1}^{(i)},h^2 \boldsymbol{V}_{t-1})$\\
    \quad\quad where $ \vc{m}_{t-1}^{(i)}$ and $\vc{V}_{t-1}$ are given in \eqref{eq:7} and \eqref{eq:9}.
    \STATE Propagate state particles $\vc{x}_t^{(i)} \sim p(\vc{x}_t|\vc{x}_{t-1}^{(i)},\boldsymbol{\theta}^{(i)})$
    \STATE Assign weights $w_t^{(i)} \propto \frac{p(\vc{y}_t|\vc{x}_{t}^{(i)},\boldsymbol{\theta}^{(i)})}{p(\vc{y}_t|\boldsymbol{\mu}_{t}^{(i)},\vc{m}_{t-1}^{(i)})}$
    \ENDFOR
  \end{algorithmic}
\end{algorithm}

\section{Adaptive Parameter Estimation}
\label{sec:adapt-param-estim}

Particle filters designed for parameter estimation, such as the Liu and West filter~\cite{West} or particle learning filter~\cite{Carvalho2009} treat the estimated parameters as strictly fixed. In most cases this means that the marginal posterior distribution of the parameters will become increasingly concentrated around a single value as more observations are observed. As a result, if the parameters are time-varying then these filters often collapse, as they are unable to adapt to any abrupt change in the parameter.

For tracking applications it is more realistic to consider time-varying parameters where the parameters change abruptly at a set of unknown time-points. For example, in Section \ref{sec:results} we shall consider the case of tracking a maneuvering target where the parameter vector which determines the target's trajectory changes depending on the target's maneuvers. This problem can be solved by bringing together changepoint models with parameter estimation methods. In order to emphasize that the parameters are no longer static but are piecewise time-varying we change the parameter notation from $\boldsymbol{\theta}$ to $\boldsymbol{\theta}_t$ which now accounts for the time index $t$.

\subsection{Changepoint Approach}
\label{sec:changepoint-analysis}

In some applications there are models whereby some of the parameters are fixed while others are time-varying. To account for such models we shall partition the parameter vector $\boldsymbol{\theta}_t$ into fixed and time-varying parameters (see Section \ref{sec:targ-track-model} for an example). This approach is advantageous for target tracking problems, where initially there may be several unknown parameters causing high variability in the state estimates. Over time this variability will decrease as the filter refines the estimate of the fixed parameters while still allowing the time-varying parameters to change according to the target's maneuvers. This approach is preferable compared to model switching schemes such as the IMM filter which handles fixed and time-varying parameters in the same manner and therefore does not benefit from fixing some subset of the parameters over time.

The fixed parameters can be estimated using the techniques outlined in Section \ref{sec:bayes-param-estim}. As for the time-varying parameters, we focus on the case where the parameters are piecewise constant through time. Thus there will be a set of unknown points in time, known as changepoints, where the parameters can change. We use segments to denote the time-periods between changepoints, with parameters are assumed to be constant within each segment.

Rather than estimate these changepoints, and then perform inference conditional on a set of inferred changepoints, we introduce a probabilistic model for the location of changepoints and perform inference by averaging over the resulting uncertainty in changepoint locations. For simplicity our prior model for changepoints is that there is a probability, $\beta$, of a changepoint at each time-point; and that changepoints occur independently. We assume that $\beta$ is known. For a given $\beta$ value, the expected segment length is $1/\beta$. Thus prior knowledge about the length of segments can be used to choose a reasonable value of $\beta$ to use for a given application. In practice the data often gives strong indication about the location of changepoints, and thus we expect the results to be robust to reasonable choices of $\beta$. This is shown empirically in a simulated example (Section \ref{sec:choice-beta}), where we observed that similar results are obtained for values of $\beta$ varying by about an order of magnitude.

If there is a changepoint at time $t$, then new parameter values will be drawn from some distribution $p_{\boldsymbol{\theta_{t-1}}}(\cdot)$ which depends on the current parameter values, $\boldsymbol{\theta_{t-1}}$. For ease of notation we consider distributions where we can partition the parameter, $\boldsymbol{\theta}=(\boldsymbol{\theta}^{\prime},\boldsymbol{\theta}^{\prime\prime})$, into components that are fixed and those which change to a value independent of the current parameter value; though more general choices of distribution are possible. Thus we assume
\begin{equation}
  \label{eq:13}
 p_{\boldsymbol{\boldsymbol{\theta}_{t-1}}}(\boldsymbol{\theta}_t) = \delta(\boldsymbol{\theta}_t^\prime-\boldsymbol{\theta}^{\prime}_{t-1}) p(\boldsymbol{\theta}^{\prime\prime}_t),  
\end{equation}
where $\delta(\cdot)$ is the Dirac-delta function, and $p(\cdot)$ is some known density function. It is natural to assume that $p(\cdot)$ corresponds to the prior distribution for $\boldsymbol{\theta}^{\prime\prime}_1$.
Thus the parameter dynamics can be described as
\begin{eqnarray}
\label{eq:12}
  \boldsymbol{\theta}_t = \left\{
  \begin{array}{l l}
    \boldsymbol{\theta}_{t-1} & \quad \text{with probability $1-\beta$,}\\
    \boldsymbol{\gamma}_t & \quad \text{with probability $\beta$,}\\
  \end{array} \right.
\end{eqnarray}
\noindent where  $\boldsymbol{\gamma}_t \sim p_{\boldsymbol{\theta}_{t-1}}(\cdot)$ represents the new parameter values.

\subsection{SMC Inference for Time-Varying Parameters}
\label{sec:smc-inference-with}

It is straightforward to implement SMC inference under our changepoint model for parameters, whereby we simulate parameter values from \eqref{eq:12} as part of the state update at each iteration. However this naive implementation can be improved upon using the ideas behind the APF filter to update the prior probability of a changepoint $\beta$ with the newest observations $\vc{y}_t$.
Consider the posterior distribution $p(\vc{x}_t,\boldsymbol{\theta}_t,k|\vc{y}_{1:t})$ which from \eqref{eq:12} now takes account of the potentially new parameter vector $\boldsymbol{\gamma}_t$. Only one of the
parameter vectors $\boldsymbol{\theta}_{t-1}$ or $\boldsymbol{\gamma}_t$ is chosen with probabilities $1-\beta$ and $\beta$, respectively. Therefore our posterior can be written as
\begin{align*}
& p(\vc{x}_t,\boldsymbol{\theta}_t,k|\vc{y}_{1:t}) \propto  \\
& (1-\beta) p(\vc{y}_t|\vc{x}_t,\boldsymbol{\theta}_{t-1}^{(k)})p(\vc{x}_t|\vc{x}_{t-1}^{(k)},\boldsymbol{\theta}_{t-1}^{(k)})\delta(\boldsymbol{\theta}_t-\boldsymbol{\theta}_{t-1}^{(k)})w_{t-1}^{(k)} \\
&+ \quad \beta p(\vc{y}_t|\vc{x}_t,\boldsymbol{\theta}_t^{(k)})p(\vc{x}_t|\vc{x}_{t-1}^{(k)},\boldsymbol{\theta}_t^{(k)})p_{\boldsymbol{\theta}^{(k)}_{t-1}}(\boldsymbol{\theta}_t^{(k)})w_{t-1}^{(k)}.
\end{align*}

Using the auxiliary particle filter outlined in Section \ref{sec:auxil-part-filt} it is possible to sample from this posterior distribution with an appropriate proposal distribution using the \textit{resample-propagate}
approach.

At time $t-1$ the posterior is represented by a set of equally-weighted particles $\{\vc{x}_{t-1}^{(i)},\boldsymbol{\theta}_{t-1}^{(i)}\}_{i=1}^N$. Each particle is given a weight proportional to its predictive likelihood, corresponding to either a changepoint or no changepoint. For $N$ particles this leads to $2N$ weights where for $i=1,\ldots,N$

\[
w_{t,1}^{(i)} \propto p(\vc{y}_t|\boldsymbol{\mu}_t^{(i)},\boldsymbol{\theta}_{t-1}^{(i)}),\mbox{ where } \boldsymbol{\mu}_t^{(i)}=\mathbb{E}[\vc{x}_{t}|\vc{x}_{t-1}^{(i)},\boldsymbol{\theta}_{t-1}^{(i)}]
\]

\noindent and

\[
w_{t,2}^{(i)} \propto p(\vc{y}_t|\boldsymbol{\mu}_t^{(i)},\boldsymbol{\gamma}_t^{(i)}) \mbox{ with } \boldsymbol{\gamma}_t^{(i)} \sim p_{\boldsymbol{\theta}^{(i)}_{t-1}}(\cdot).
\]
The first of these weights is an estimate of the probability of $\vc{y}_t$ given the value of the $i$th particle at time $t-1$ when there is no changepoint. The second weight corresponds to there being a changepoint
with new parameters $\boldsymbol{\gamma}_t^{(i)}$.

Next, resampling is performed, where $N$ particles are sampled from $2N$ particles with probabilities proportional to the union of $\{(1-\beta) w_{t,1}^{(i)}\}_{i=1}^N$ and $\{\beta w_{t,2}^{(i)}\}_{i=1}^N$. If the $i$th index is sampled from the first set of the union, then the particle corresponding to the state and current parameters for index $i$ are propagated. If the $i$th index is sampled from the second set, the state of the corresponding particle is propagated together with the new parameter value, $\boldsymbol{\gamma}_t^{(i)}$. Finally, the appropriate weights for the particles are calculated as in the auxiliary
particle filter.

Within this approach it is possible to use either the particle learning filter (Algorithm \ref{alg1}), the Liu and West filter (Algorithm \ref{alg2}) or both to update the parameter values in the segments between changepoints. The parameter vector can be partitioned as follows $\boldsymbol{\theta}_t =(\boldsymbol{\xi}^\top_t,\boldsymbol{\zeta}^\top_t)^\top$ where $\boldsymbol{\xi}_t$ are parameters to be updated using the particle learning filter and $\boldsymbol{\zeta}_t$ are parameters updated using the Liu and West filter. This is a slight abuse of notation as $\boldsymbol{\theta}_t$ is further partitioned into fixed and time-varying parameters. It is possible to resolve this problem by partitioning $\boldsymbol{\xi}_t$ and $\boldsymbol{\zeta}_t$ into fixed and time-varying parameters. 

\subsection{Applying the Liu and West Filter to Time-Varying Parameters}
\label{sec:applying-liu-west}

At time $t-1$ parameters $\boldsymbol{\zeta}_{t-1}$ with no sufficient statistic structure can be updated with the Liu and West filter by first estimating the kernel locations
$\vc{m}_{t-1}^{(i)}=a \boldsymbol{\zeta}_{t-1}^{(i)}+(1-a)\overline{\boldsymbol{\zeta}}_{t-1}$, where $a$ is the shrinkage parameter. The $i$th kernel location is propagated and the
parameters are updated as $\boldsymbol{\zeta}_t^{(i)} \sim \mathcal{N}(\cdot|\vc{m}_{t-1}^{(i)},h^2 \mt{V}_{t-1})$ if the index $i \in \{1,\ldots,N\}$, where $\mt{V}_{t-1}$ is given in \eqref{eq:9}.
Alternatively, if $i \in \{N+1,\ldots,2N\}$ then $\boldsymbol{\zeta}_t^{(i)}$ is drawn from the appropriate part of distribution \eqref{eq:13}.

\subsection{Applying Particle Learning to Time-Varying Parameters}
\label{sec:using-part-learn}

The particle learning filter can be viewed as a special case of the Bayesian parameter estimation approach where the parameters $\boldsymbol{\xi}_t$ have a conjugate prior distribution which can be recursively updated via the sufficient statistics $\vc{s}_{t}$. The sufficient statistics are updated differently depending on whether the parameters are fixed or time-varying. For the case of the fixed parameters the sufficient statistics are updated as described in Section \ref{sec:particle-learning}, where $\vc{s}_{t}^{(i)} = \mathcal{S}(\vc{s}_{t-1}^{(i)},\vc{x}_t^{(i)},\vc{y}_t)$ for $i \in \{1,\ldots,2N\}$.

 If we assume that $\boldsymbol{\xi}_t$ is a time-varying parameter then the parameters are updated at time $t$ by sampling $\boldsymbol{\xi}_{t}^{(i)} \sim p(\cdot|\vc{s}_{t}^{(i)})$, where $\vc{s}_{t}^{(i)} = \mathcal{S}(\vc{s}_{t-1}^{(i)},\vc{x}_t^{(i)},\vc{y}_t)$ if no changepoint is detected (i.e. the resampling index $i \in \{1,\ldots,N\}$). Alternatively, if there is a changepoint and $i \in \{N+1,\ldots,2N\}$ then the sufficient statistics are reset to their initial prior values, $\vc{s}_{t-1} = \vc{s}_{0}$ (see Section \ref{sec:targ-track-model} for an example). If some parameters are fixed and others time-varying then the sufficient statistics for each parameter are updated accordingly.

Applying the Liu and West filter and the particle learning filter to the estimation of time-varying parameters produces an efficient filter for both state and parameter estimation which we refer to as the adaptive parameter estimation (APE) filter. Algorithm \ref{alg4} presents an instance of the filter where the parameters $\boldsymbol{\zeta}_{t}$ are assumed to be time-varying and the parameters $\boldsymbol{\xi}_t$ are assumed to be fixed. This setting conforms with the scenario given in the performance validation section.

\begin{algorithm}
\caption{Adaptive Parameter Estimation Filter for Fixed and Time-Varying Parameters}
\label{alg4}

  \begin{algorithmic}
    \FOR{$i =1,\ldots,N$}
    \STATE Update parameter values: \\
    \quad  $\boldsymbol{\xi}_{t-1}^{(i)} \sim p(\cdot|\vc{s}_{t-1}^{(i)})$ \\
    \quad  $\vc{m}_{t-1}^{(i)}=a \boldsymbol{\zeta}_{t-1}^{(i)}+(1-a)\overline{\boldsymbol{\zeta}}_{t-1}$ \\
    \quad  Set parameter vector $\boldsymbol{\theta}_{t}^{(i)} = [{\boldsymbol{\xi}_{t-1}^{(i)}}^\top,{\vc{m}_{t-1}^{(i)}}^\top]^\top$ \\
    \STATE Calculate pre-weights \\
    $w_{t,1}^{(i)}  \propto p(\vc{y}_t|\boldsymbol{\mu}_t^{(i)},\boldsymbol{\theta}_{t}^{(i)})$ where $\boldsymbol{\mu}_t^{(i)}=\mathbb{E}[\vc{x}_{t}|\vc{x}_{t-1}^{(i)},\boldsymbol{\theta}_{t-1}^{(i)}]$.\\
    \ENDFOR
    \FOR{$i =1,\ldots,N$}
    \STATE Sample new parameter particles $\boldsymbol{\gamma}_{t}^{(i)} \sim p_{\boldsymbol{\theta}_{t-1}^{(i)}}(\cdot)$
    \STATE Calculate pre-weights \\
     \quad \quad $w_{t,2}^{(i)} \propto p(\vc{y}_t|\boldsymbol{\mu}_t^{(i)},\boldsymbol{\gamma}_{t}^{(i)})$ \\
    \ENDFOR
     \FOR {$i=1,\ldots,N$}
    \STATE Sample indices $k^i$ from $\{1,\ldots,2N\}$ with \\
    probabilities $\{(1-\beta) w_{t,1}^{(i)}\}_{i=1}^{N}$ and $\{\beta w_{t,2}^{(i)}\}_{i=N+1}^{2N}$.
    \ENDFOR
    \FOR {$k^i \in \{1,\ldots,N\}$}
    \STATE Update parameters $\boldsymbol{\zeta}_t^{(i)} \sim \mathcal{N}(\cdot|\vc{m}_{t-1}^{(k^i)},h^2 \mt{V}_{t-1})$
    \STATE where $\mt{V}_{t-1}$ is given in \eqref{eq:9}.
    \STATE Set parameters $\boldsymbol{\theta}_{t}^{(i)} = [{\boldsymbol{\xi}_{t-1}^{(k^i)}}^\top,{\boldsymbol{\zeta}_{t}^{(i)}}^\top]^\top$
    \STATE and sufficient statistics $\vc{s}_{t-1}^{(i)} = \vc{s}_{t-1}^{(k^i)}$
    \STATE Propagate states $\vc{x}_t^{(i)} \sim p(\vc{x}_t|\vc{x}_{t-1}^{(k^i)},\boldsymbol{\theta}_t^{(i)})$
    \STATE Assign weights $w_t^{(i)} \propto \frac{p(\vc{y}_t|\vc{x}_{t}^{(i)},\boldsymbol{\theta}_t^{(i)})}{w_{t,1}^{(k^i)}}$
    \ENDFOR
    \FOR {$k^i \in \{N+1,\ldots,2N\}$}
    \STATE Propagate states $\vc{x}_t^{(i)} \sim p(\vc{x}_t|\vc{x}_{t-1}^{(k^i)},\boldsymbol{\gamma}_t^{(k^i)})$
    \STATE Set parameters $\boldsymbol{\theta}_{t}^{(i)} = \boldsymbol{\gamma}_{t}^{(k^i)}$
    \STATE Assign weights $w_t^{(i)} \propto \frac{p(\vc{y}_t|\vc{x}_{t}^{(i)},\boldsymbol{\theta}_t^{(i)})}{w_{t,2}^{(k^i)}}$
    \ENDFOR
    \STATE Resample particles $\{\vc{x}_t^{(i)},\vc{s}_{t-1}^{(i)},\boldsymbol{\zeta}_t^{(i)}\}_{i=1}^N$ with replacement\\
    \quad\quad\quad with probabilities $\{w_t^{(i)}\}_{i=1}^N$ to obtain the particle set\\
    \quad\quad\quad  $\{\vc{x}_t^{(i)},\vc{s}_{t-1}^{(i)},\boldsymbol{\zeta}_t^{(i)}\}_{i=1}^N$ with weights $1/N$.
    \STATE Update sufficient statistics $\vc{s}_{t}^{(i)} = \mathcal{S}(\vc{s}_{t-1}^{(i)},\vc{x}_t^{(i)},\vc{y}_t)$.
  \end{algorithmic}
\end{algorithm}

\subsection{Target Tracking Motion and Observation Models}
\label{sec:targ-track-model}

We present a motivating example from the target tracking literature to highlight the importance of estimating time-varying parameters. The model considered is used to track a target which moves within the $x-y$
plane, where the target's state is a vector of position and velocity $\vc{x}_t=(x_t,\dot{x}_t,y_t,\dot{y}_t)^\top$.

The motion of the target is modeled using a coordinated-turn model \cite{Rong1993} of the form
\begin{equation*}
\label{eq:8}
  \mathbf{\vc{x}_t}= \mt{F}^\top{\vc{x}_{t-1}} + \boldsymbol{\Gamma}\boldsymbol{\nu}_t
\end{equation*}
where,
\begin{align*}
\mt{F} = \left[
    \begin{array}{cccc}
      1& \frac{\sin \omega_{t}\Delta T}{\omega_t} &0& -\frac{1- \cos \omega_{t}\Delta T}{\omega_{t}} \\
      0&\cos \omega_{t}\Delta T &0&- \sin \omega_{t}\Delta T  \\
      0&\frac{1-\cos \omega_{t} \Delta T}{\omega_{t}}&1&\frac{\sin \omega_{t} \Delta T}{\omega_{t}} \\
      0&\sin \omega_{t} \Delta T&0&\cos \omega_{t} \Delta T
    \end{array} \right],
\end{align*}
\begin{align*}
\boldsymbol{\Gamma} = \left[
    \begin{array}{cc}
      \frac{\Delta T}{2}& 0 \\
      \Delta T& 0  \\
      0& \frac{\Delta T}{2} \\
      0& \Delta T
    \end{array} \right]
\end{align*}
and system noise $\boldsymbol{\nu}_t$ is modeled as a zero mean Gaussian white noise process $\mathcal{N}(0,\eta^2 I_2)$.

This model simplifies to the constant velocity model when $\omega_t=0$. The model is flexible and able to account for the motion of highly maneuverable targets, where the target may change direction abruptly and switch between periods of high and low maneuverability (see Figure \ref{fig:track} for a simulated trajectory).

Noisy nonlinear observations of the target in the form of a range and bearing measurement are taken by a fixed observer positioned at $(s_x,s_y)$
\begin{align*}
  \mathbf{\mt{y}_t}= \left[
    \begin{array}{c}
      \sqrt{(x_t-s_x)^2+(y_t-s_y)^2} \\
      \arctan((y_t-s_y)/(x_t-s_x))
    \end{array} \right] + \boldsymbol{\epsilon}_t,
\end{align*}
where the observation noise $\boldsymbol{\epsilon}_t$ is a zero mean Gaussian white noise process with known covariance matrix $\mt{R}$.

It is possible to use this model to track a maneuvering target if we treat the turn rate parameter $\omega_t$ as a time-varying parameter and the remaining parameters $\eta^2$ and $\mt{R}$ as fixed. This is an ideal scenario for the adaptive parameter estimation filter as it can easily handle both fixed and time varying parameters. In Section \ref{sec:results} a comparison of this filter with the IMM filter illustrates the benefit of treating fixed and time-varying parameters separately.

The APE filter can be applied to the target tracking model in the following way. The turn rate parameter $\omega_t$ appears non-linearly in the model and does not admit a sufficient statistic structure. We therefore estimate this parameter using the kernel density approach for time-varying parameters as outlined in Section \ref{sec:applying-liu-west}. The noise variance parameters $\eta^2$ and $\mt{R}$ can be estimated via the set of sufficient statistics $\vc{s}_t=(a_t,b_t,c_t,d_t,e_t,f_t)$ which is a vector of the parameters for the conjugate priors. The conjugate prior for $\eta^2$ is an inverse-gamma distribution $IG(a_t/2,b_t/2)$, where the sufficient statistics $a_t$ and $b_t$ are updated as $a_t=a_{t-1}+\mbox{dim}(\mt{x}_t)$ and $b_t = b_{t-1} + (\mt{x}_t-\mt{F}^\top\mt{x}_{t-1})^\top(\mbox{diag}(\boldsymbol{\Gamma}\boldsymbol{\Gamma}^\top))^{-1}(\mt{x}_t-\mt{F}^\top\mt{x}_{t-1})$. The conjugate prior for $\mt{R}$ is an inverse Wishart distribution. However, if we assume that the range and bearing measurements are uncorrelated then we can model their variances separately, where the range variance follows an inverse-gamma distribution $IG(c_t/2,d_t/2)$, with the sufficient statistics $c_t$ and $d_t$ which are updated as follows, $c_t=c_{t-1}+1$ and $d_t=d_{t-1}+(\mt{y}_{t}[1]-\sqrt{(\mt{x}_t[1]-s_x)^2+(\mt{x}_t[3]-s_y)^2})^2$ and the sufficient statistics $e_t$ and $f_t$ for the variance of the bearing measurements are updated similarly. 

The example given in Section \ref{sec:results} treats the variances as fixed and therefore we do not need to reset the sufficient statistics for these parameters when a changepoint is detected. It is possible to allow one of the variances to change between segments by resetting a subset of the sufficient statistics. For example, it may be reasonable to assume that the variance of observations $\mt{R}$ is fixed, but that $\nu^2$ changes when the target performs a maneuver. The change in $\nu^2$ can be accounted for by setting $a_t=a_0$ and $b_t=b_0$, thus the new variance parameter will be sampled from the initial prior distribution. The sufficient statistics will again be updated accordingly to estimate $\nu^2$ as given above.

To summarize, the adaptive parameter estimation filter can be used to estimate fixed and time-varying parameters for models with both conjugate and non-conjugate parameter distributions.
In the next section we will show how this approach works well when there are multiple unknown parameters with vague prior knowledge of their true values.

\section{Performance Validation}
\label{sec:results}

This section presents a comparison of the adaptive parameter estimation filter developed in Section~\ref{sec:adapt-param-estim} against the IMM filter. The filters' performance is validated on a simulated dataset taken from the coordinated turn model given in Section \ref{sec:targ-track-model}. The aim of the comparisons is to illustrate the improvement of the APE filter over the IMM filter as the number of unknown parameters increases. The accuracy of the algorithms is characterized by the relative root mean squared (RMS) representing the ratio, i.e. the IMM RMS error/APE RMS error. Results showing the filters' accuracy and computational time are given.

\subsection{Testing Scenario}

A challenging testing scenario is considered in which the moving object performs complex maneuvers consisting of abrupt turns followed by a straight line motion. The turn rate parameter $\omega_{t} \in (-20^\circ/s,20^\circ/s)$ is unknown and is estimated in conjunction with the target's state vector. For variance parameters $\nu^2$ and $\mt{R}$ the initial parameters for the conjugate priors are $\vc{s}_0=(9,15,4,5000,4,0.0025)$. A target track is simulated from the coordinated turn model over 400 time steps, with sampling period $\Delta T=1s$. The turn rate parameter $\omega_{t}$ takes values $\{0,3,0,5.6,0,8.6,0,-7.25,0,7.25\}^\circ/s$ with changes occurring at times $\{60,120,150,214,240,272,300,338,360\}$, respectively. This set-up creates a highly dynamic target trajectory, where the target switches between periods of high and low maneuverability, as shown in Figure \ref{fig:track}. The testing scenario is completed by specifying the system noise variance $\eta^2=2 \mbox{m/s}^2$ and observation noise covariance matrix $\mt{R}=\mbox{diag}(50^2\mbox{m},1^{\circ})$. The trajectory is simulated with the initial state of the target $\mt{x}_1 = (30 \mbox{km},300 \mbox{m/s},30 \mbox{km},0 \mbox{m/s})^\top$ and observations taken from a fixed observer positioned at $(55 \mbox{ km},55 \mbox{km})$.

\subsection{Choosing $\beta$}
\label{sec:choice-beta}

\begin{figure}[t]
  \centering
  \includegraphics[width=2.5in]{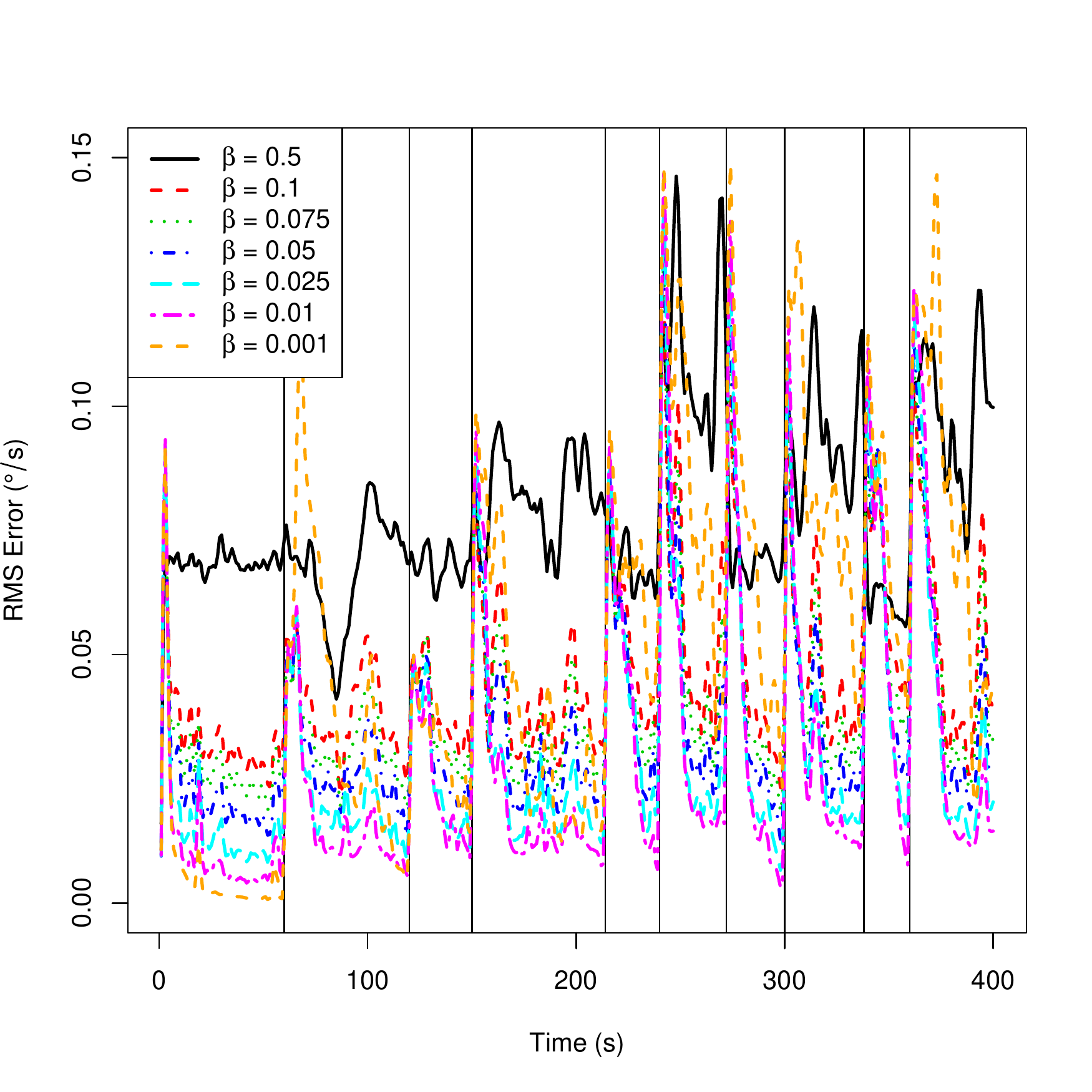}
  \caption{Root mean squared of the turn rate parameter from model \eqref{eq:8} for various $\beta$ values.}
  \label{fig:rmse_turn_rate}
\end{figure}

The accuracy of the APE filter is dependent on the choice of the \textit{a priori} changepoint probability $\beta$. If $\beta$ is large (close to 1) then the filter may struggle to estimate the parameters as it will introduce excess parameters from the  diffuse prior $p_{\boldsymbol{\theta}_1}(\boldsymbol{\gamma}_t)$ when no changepoint has occurred. On the other hand, if $\beta$ is too small then the filter will simplify to the standard Bayesian parameter estimation filter for static parameters, and will struggle to handle time-varying parameters. Figure \ref{fig:rmse_turn_rate} gives the RMS error for the parameter $\omega_t$ using the APE filter with various choice for $\beta$ using the simulated trajectory described in Section \ref{sec:example-1.-estim}. The vertical lines correspond to the changepoints where the target performs a maneuver. In this scenario there are 9 changepoints over 400 time steps, therefore using the inverse of the average segment length we would expect $\beta \approx 0.025$ to give the lowest RMS error. The results show that setting $0.01<\beta<0.05$ will give the lowest RMS error, consistent with results from other simulated trajectories. The filter does not require that the changepoint probability $\beta$ parameter is known exactly. In fact the filter appears to be robust to a range of $\beta$ values. For example, when $\beta=0.001$ the filter displays higher RMS error after a changepoint, this is to be expected as setting $\beta$ close to $0$ assumes there is no changepoint. However, even for such low values the filter is still able to track the target. This is in contrast to the Liu and West and particle learning filters which often collapse when used to estimate abruptly changing parameters (see Figure \ref{fig:track}).

\subsection{Estimation of the state vector, jointly with the turn rate.}
\label{sec:example-1.-estim}

The APE filter is compared with the IMM filter to estimate the state
vector of a maneuvering target. The main difference between these
two approaches is in the way that each filter handles the unknown
turn rate $\omega_t$. The IMM attempts to account for the unknown,
time-varying turn rate by selecting one model from a bank of
potential models. The adaptive parameter estimation filter, on the
other hand, estimates $\omega_t$ and is therefore not constrained by
a finite set of potential models.
\begin{figure}[!t]
  \centering
  \includegraphics[width=2.5in]{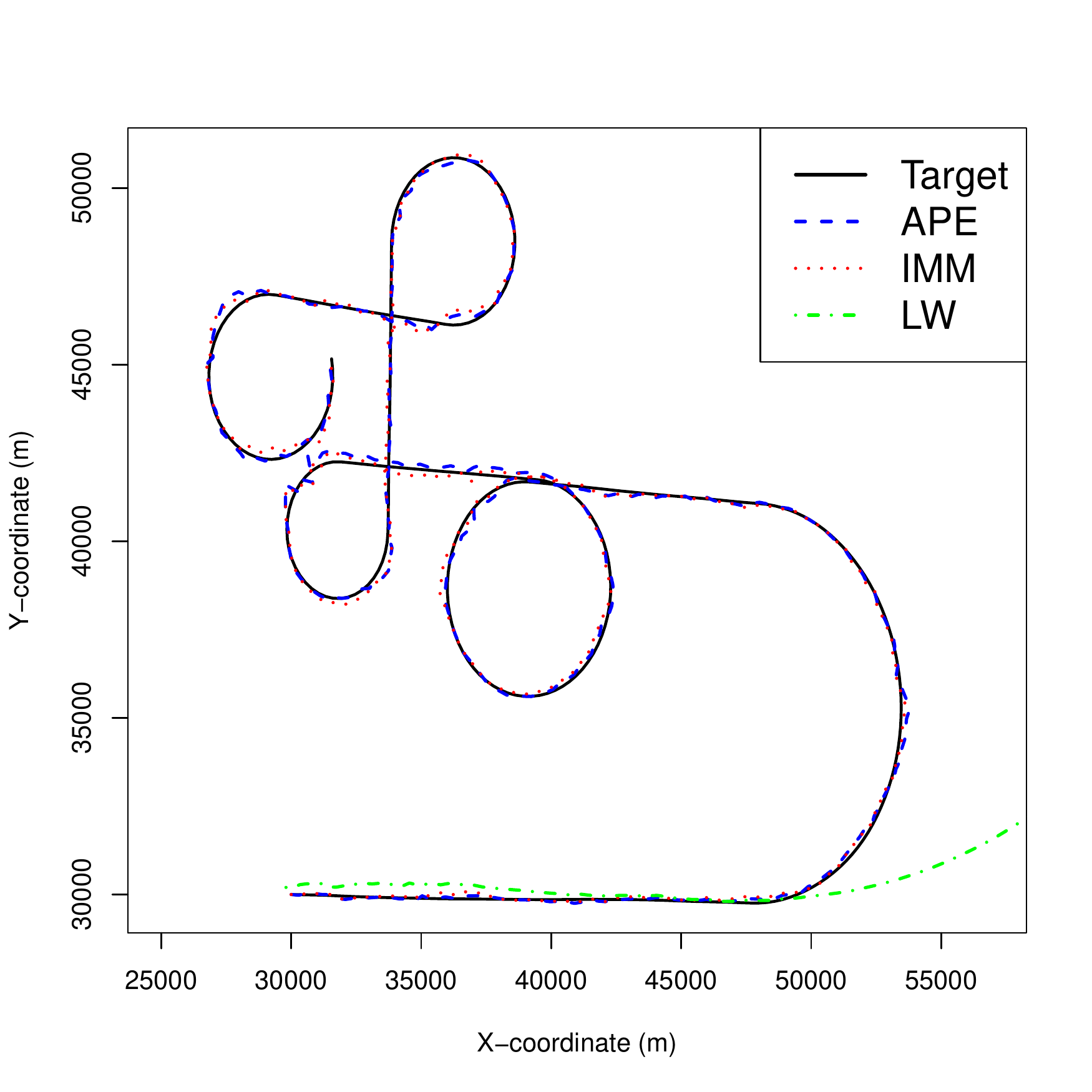}
  \caption{This Figure shows the simulated target trajectory and the estimated trajectories obtained by the APF, IMM and LW algorithms. }
  \label{fig:track}
\end{figure}

The APE filter is implemented with 5,000 particles and as there does
not exist a conjugate prior for the turn rate parameter $\omega_t$,
the Liu and West procedure shall be used within Algorithm \ref{alg4}
to estimate this parameter. The smoothing parameter of the kernel
density estimate is set to $h^2=0.01$ as recommended \cite{West},
where the probability of a changepoint at any point in time is
$\beta=0.05$. The initial prior distribution for the turn rate
parameter $\omega_{t}$ follows a non-informative uniform
distribution over the range $[-20^\circ, 20^\circ]$. In this
scenario the IMM filter is implemented using 20 and 60 coordinated
turn models. The models differ in the choice of the parameters
$\omega_{t}$ and $\eta^2$, where 20 or 60 equally spaced values of
$\omega_{t}$ are sampled over the range $[-20^\circ, 20^\circ]$ and
$\eta^2 = 2 \mbox{m/s}^2$ when $\omega_t=0$ and $2.5 \mbox{m/s}^2$
when the turn rate is non-zero to allow for greater ease of turn.

The transition probabilities between models of the IMM filter are
balanced equally between all alternative models and sum to 0.05 with
a 0.95 probability of no model transition. This parameter acts in a
similar way to the $\beta$ parameter of the APE filter and must also
be tuned. For this example we have set the model transition
probability to be equal to $\beta$ to create a fair comparison. As
the observation model is nonlinear the IMM filter is implemented
with an unscented Kalman filter~\cite{julieranduhlmann:2004:ufane}.
In this setting the computational time required to run the IMM
filters, relative to the adaptive parameter estimation filter, is
0.5 and 2.5 times greater for 20 and 60 models respectively. The
filters are compared over 100 independent Monte Carlo runs.

\begin{figure}[!t]
  \centering
  \includegraphics[width=2.5in]{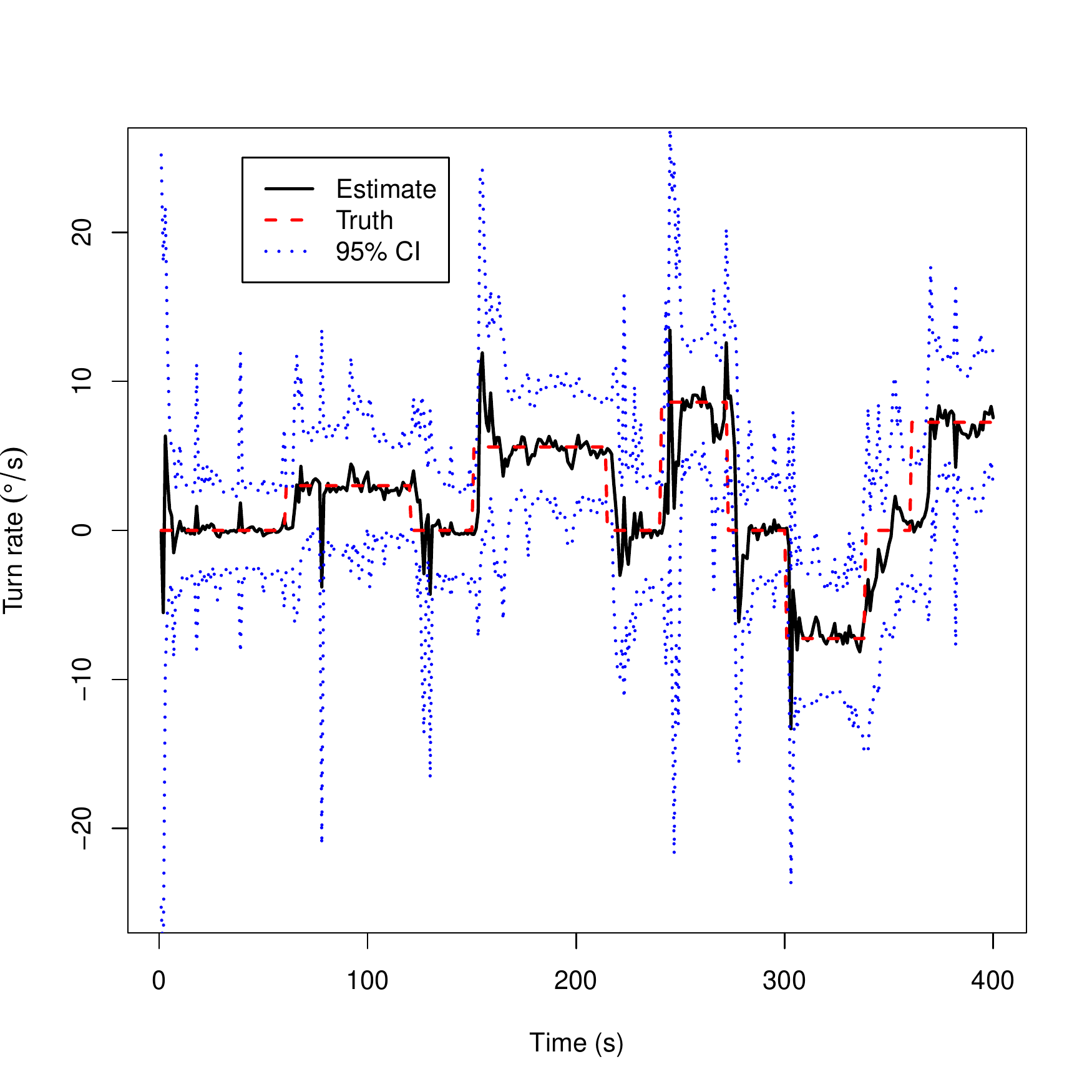}
  \caption{Estimated turn rate parameter (black solid line) with the APE filter versus the true parameter value (red dashed line)}
  \label{fig:turn_rate}
\end{figure}

Simulation results show that both the IMM and the adaptive parameter
estimation filter are able to track the target well. However, if the
standard Liu and West (LW) filter (Algorithm \ref{alg2}) with no
adaptation is applied to this scenario then after the first
maneuver, when the turn rate changes, the parameter estimated by the
LW filter no longer matches the target's dynamics, which after a few
time steps causes the filter to collapse (Fig. \ref{fig:track}). The
efficacy of the adaptive parameter estimation filter is dependent
upon the accuracy of the parameter estimates. Figure
\ref{fig:turn_rate} shows the estimates of the unknown turn rate
given by the APE filter. The filter appears to estimate the turn
rate well under difficult conditions. During long periods between
maneuvers the filter is able to produce reliable estimates of the
turn rate parameter and update this estimate to account for changes
in the target's dynamics.

\begin{figure*}[!t]
  \centerline{
  \subfigure{\includegraphics[width=3in]{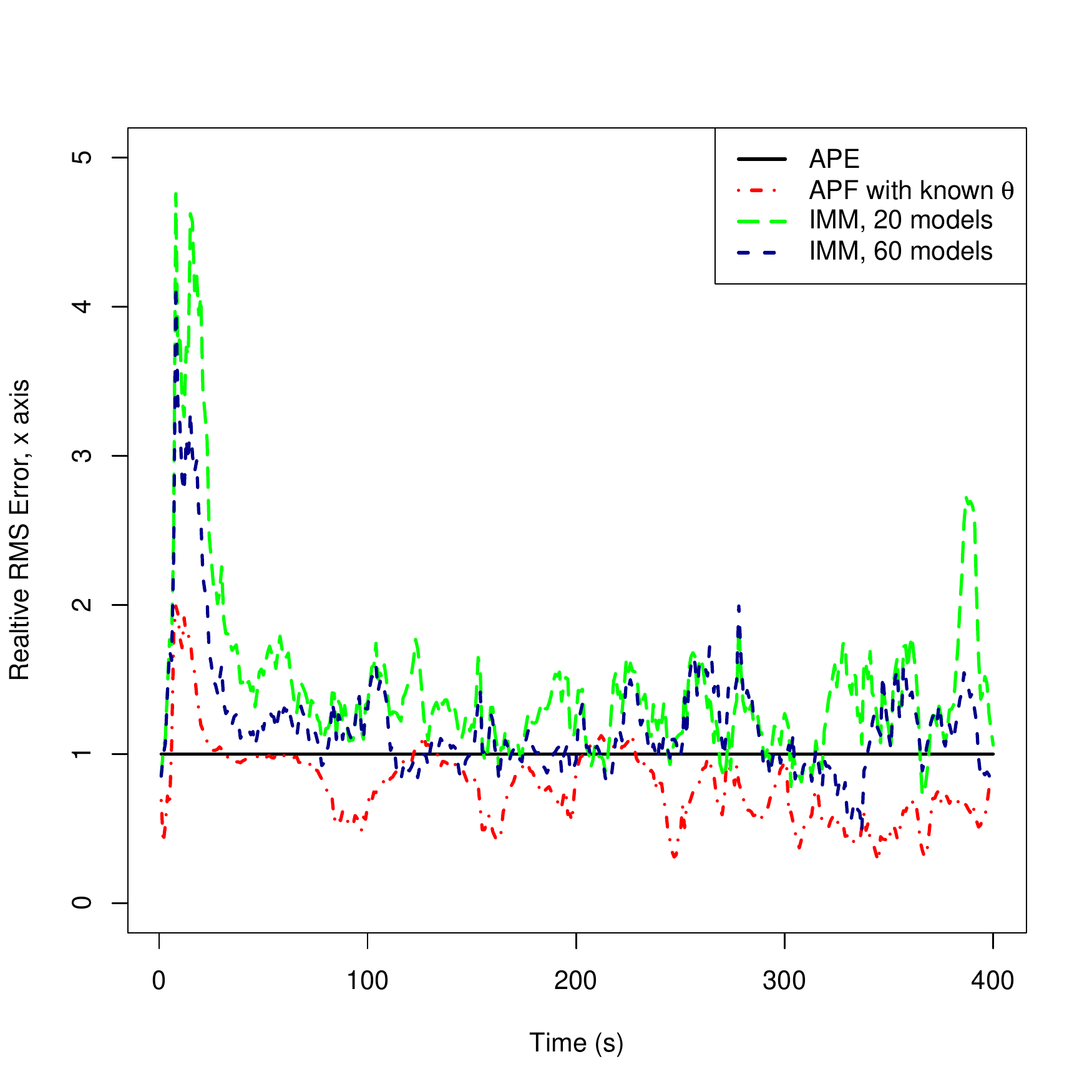}}
   \subfigure{\includegraphics[width=3in]{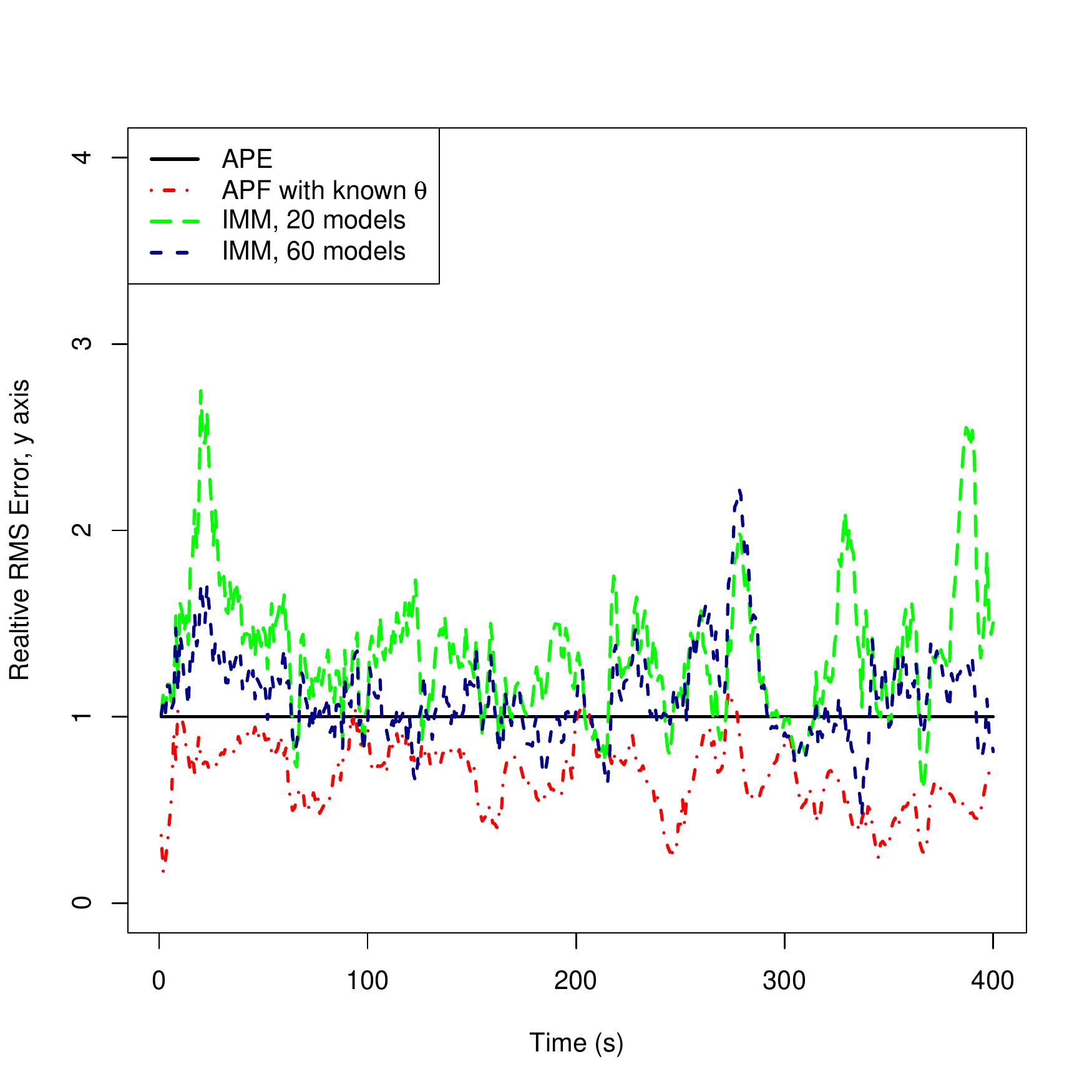}}}
   \caption{Relative RMS error (IMM RMS error/APE RMS error) of target position for the $x$ and $y$ axes, respectively (top $x$ axis, bottom $y$ axis.)}
  \label{fig:rmsePV}
\end{figure*}

Figure \ref{fig:rmsePV} gives the RMS error of several filters
relative to the APE filter. It also displays a comparison to the auxiliary particle filter where $\theta$ is known. This comparison illustrates the importance and potential gains that are achievable by correctly estimating the unknown model parameters. Improvements in the accuracy of the IMM filter may be attained by tuning the filter to better match the dynamics of the target. However, with minimal tuning, the adaptive parameter estimation filter is able to track the target at least as well as the IMM and requires no prior knowledge of the target's dynamics.

The average RMS error over the
trajectory for the following filters: ``APE'', ``IMM 20 models'',
``IMM 60 models'' and ``APF filters'' is, 81.41m, 110.23m, 92.97m
and 61.86m, respectively. Compared to both IMM filters the APE
filter produces lower RMS error of the target's position. The
benefit of the APE filter is most notable during longer segments
between changepoints. This is to be expected as longer segments
allow the APE filter to refine its estimate of the turn rate
parameter. Increasing the number of models for the IMM filter can
reduce the RMS error, but at an increase in computational
complexity. In the next section we shall see that, computational
complexity aside, increasing the number of models does not guarantee
a reduction in RMS error.

\subsection{Estimation of the state vector, jointly with the turn rate, system and observation covariance parameters.}
\label{sec:example-2.-estim}

\begin{figure*}[!t]
  \centerline{
  \subfigure{\includegraphics[width=3in]{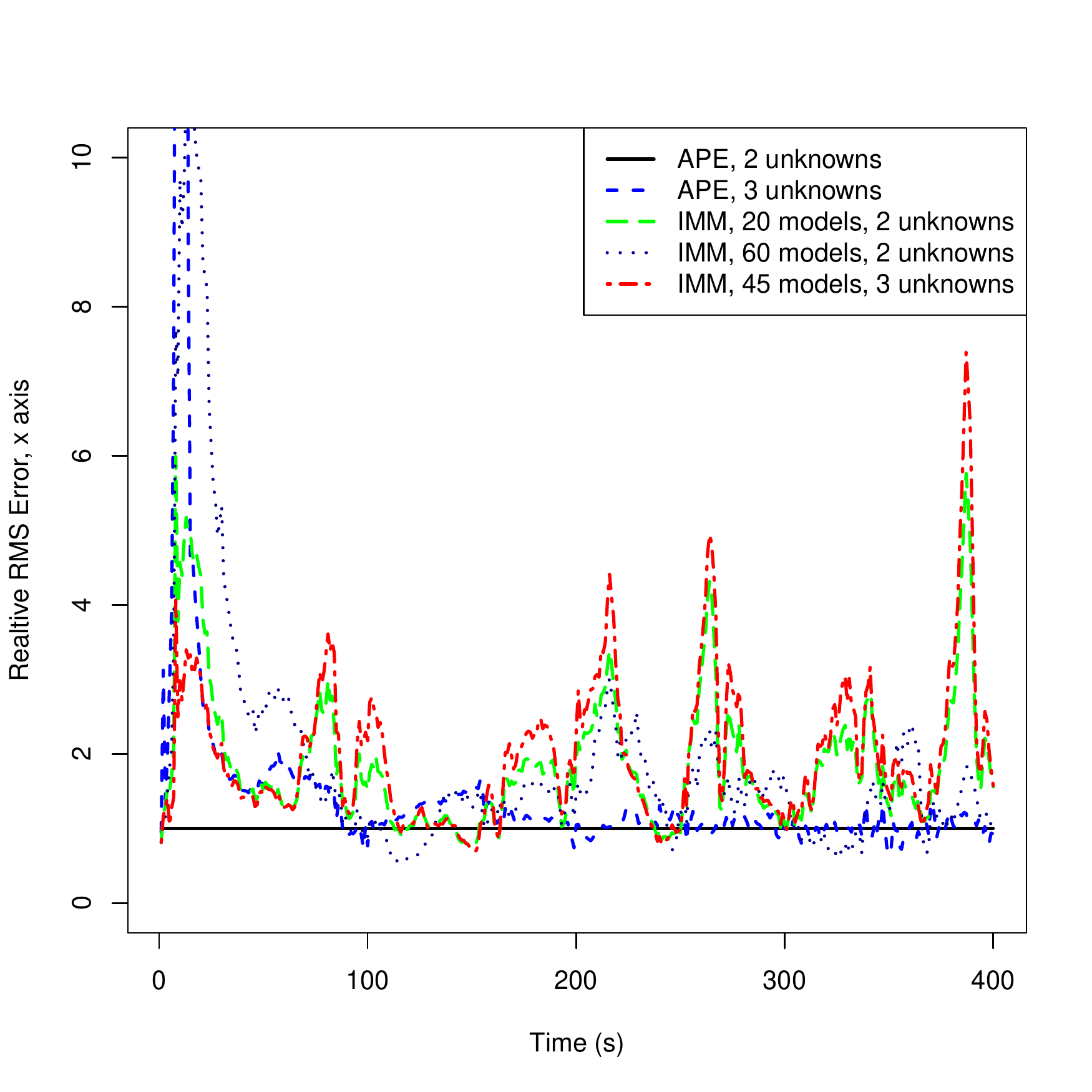}}
   \subfigure{\includegraphics[width=3in]{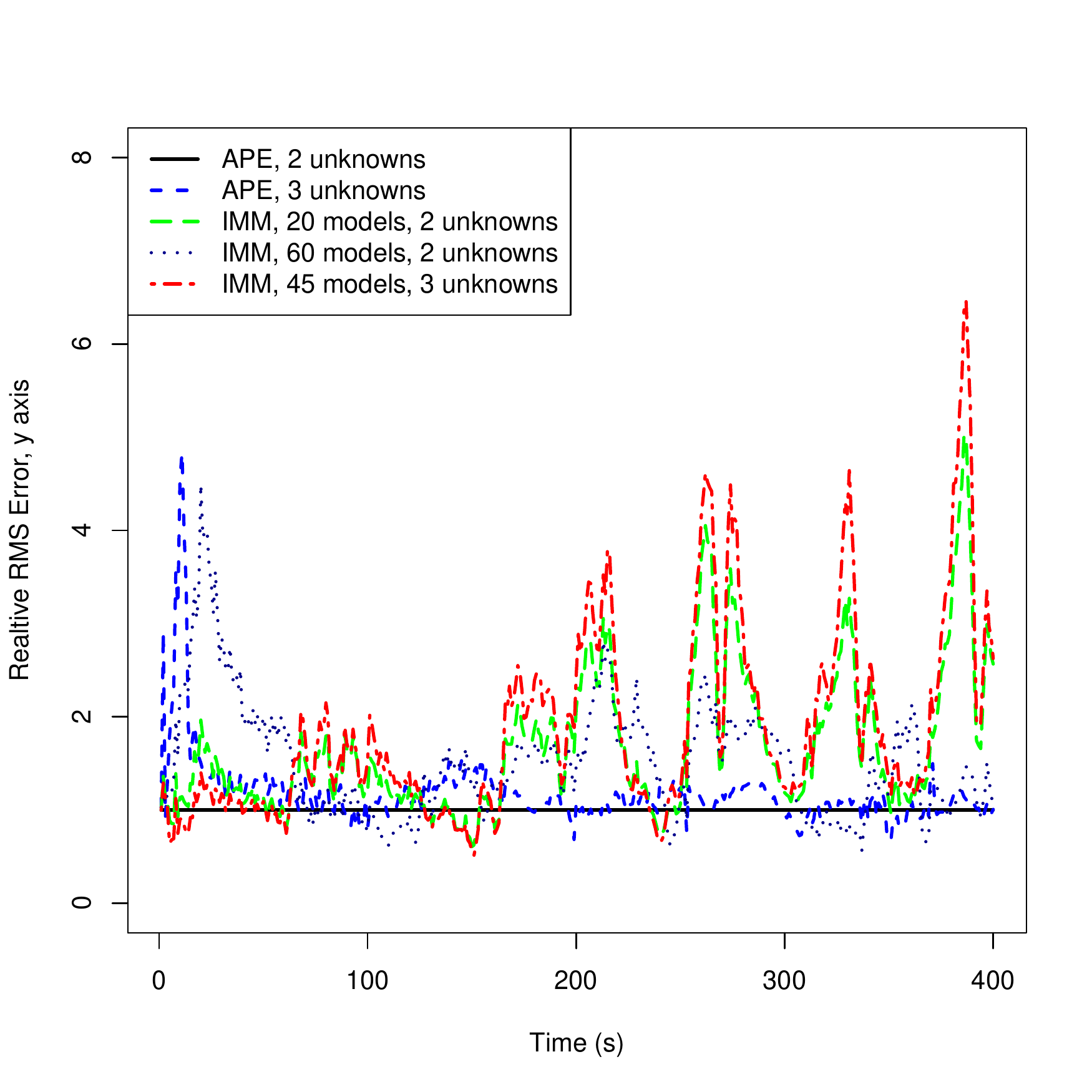}}}
   \caption{Relative RMS error of target position with multiple unknown parameters (top $x$ axis, bottom $y$ axis.)}
  \label{fig:rmsePV2}
\end{figure*}

In scenarios where there are multiple unknown parameters it becomes increasingly difficult to design an effective IMM filter as increasing the number of unknown parameters requires an increase in the number of
potential model combinations. Figure \ref{fig:rmsePV2} displays the RMS error of the APE and IMM filters when the turn rate $\omega_t$, system noise $\eta^2$ and observation covariance $\mt{R}$ parameters are unknown.
This is an interesting problem as the turn rate parameter is treated as piecewise time-varying and the variances of the noise parameters are assumed to be fixed. In this setting the APE filter uses the Liu and West
filter to estimate the turn rate parameter as in the last example and uses the particle learning filter to estimate the noise variances via their sufficient statistics (see Section \ref{sec:targ-track-model} for details).

Implementing the IMM filter becomes more complicated as the number of unknown parameters increases and the number of model combinations will also increase. If we assume that only $\omega_t$ and $\eta^2$ are unknown
then one potential implementation of the IMM filter with 20 models would be $\omega_t \in [-20^\circ/s,-10^\circ/s,0^\circ/s,10^\circ/s,20^\circ/s]$ and
$\eta^2 \in [1.5\mbox{m/s}^2,2\mbox{m/s}^2,2.5\mbox{m/s}^2,3\mbox{m/s}^2]$. Or using 60 models it would be possible to have 10 models for $\omega_t$ evenly sampled from the interval $[-20^\circ/s,20^\circ/s]$ and
6 models for $\eta^2$. This quickly leads to a combinatorial problem where it becomes difficult to match the various model combinations required to cover the unknown parameters. Increasing the number of models from
20 to 60 incurs a 3 fold increase in computational time but only offers marginal increase in the number of model combinations.

Figure \ref{fig:rmsePV2} gives the RMS error for the APE and IMM filters when 2 parameters are unknown $\{\omega_t,\eta^2\}$ and when 3 parameters are unknown $\{\omega_t,\eta^2,\mt{R}\}$. The RMS error is plotted
relative to the APE filter for 2 unknown parameters. For the case of 3 unknown parameters the IMM is implemented with 45 models combined from: 5 models for $\omega_t$ evenly sampled from the interval
$[-20^\circ/s,20^\circ/s]$, 3 models for $\eta^2 \in [2\mbox{m/s}^2,2.5\mbox{m/s}^2,3\mbox{m/s}^2]$ and 3 models for
$\mt{R} \in [\mbox{diag}(50^2 \mbox{m},1^{\circ}),\mbox{diag}(25^2 \mbox{m},2^{\circ}),\mbox{diag}(100^2 \mbox{m},1^{\circ})]$. The average RMS error for the following filters:
``APE 2 unknowns'', ``APE 3 unknowns'', ``IMM models 20 2 unknowns'', ``IMM 60 2 unknowns'' and the ``IMM 45 models 3 unknowns'' over the trajectory is, 82.79m, 101.63m, 138.93m, 127.33m and 155.65m,
respectively. For the case of 2 unknown parameters, Figure \ref{fig:rmsePV2} illustrates that increasing the number
of models in the IMM filter (from 20 to 60) does not greatly improve state estimation given the significant increase in computational time. In this scenario as the target is initially moving with almost constant
velocity, the extra turn rate combinations are redundant, but once the target begins to maneuver the benefit of extra models is observed.

 It is important to note that for the IMM filters the true noise variances are included as potential models, whereas for the APE filter, all of the parameters are truly unknown as the initial parameter values are sampled from their prior distributions. This explains why initially the RMS error of the APE filter with 3 unknown parameters is high in Figure \ref{fig:rmsePV2}. Interestingly, the RMS error of the APE filter for 3 unknown parameters approaches the levels observed for the case of 2 unknown parameters as the parameter estimates converge to their true values. This is not the case for the IMM filter as increasing the number of unknown parameters corresponds to a consistent increase in RMS error throughout time.

\section{Conclusions}
\label{sec:conclusion}

This paper considers the difficult problem of joint state and parameter estimation of nonlinear and highly dynamic systems. The paper presents a sequential Monte Carlo filter that is capable of estimating parameters with conjugate and non-conjugate structures, but most importantly, parameters which may be time-varying as in the case of tracking maneuvering targets. The main advantage of the adaptive parameter estimation approach is its ability to provide quick estimation of the abruptly changing parameters from non-informative prior knowledge, and to do this for multiple unknown parameters. Its scalability to the case of estimating multiple unknown parameters is an advantage over filters such as the IMM which are based on a multiple model implementation.

One of the drawbacks of the particle learning approach~\cite{Carvalho2009} is the requirement that the parameters follow a conjugate structure for the sufficient statistics. This limits the class of models to which particle learning~\cite{Carvalho2009} can be applied. Recent work on the extended parameter filter \cite{Russell2013} aims to overcome this problem by considering a Taylor series approximation to the parameters. A possible extension for future work would be to apply the extended parameter filter within the adaptive parameter estimation framework.

\section*{Acknowledgements}
The authors would like to acknowledge the support from EPSRC (grant EP/G501513/1) and MBDA, UK.


\bibliography{Mila,library_J}

\begin{thebibliography}{10}
\providecommand{\url}[1]{#1}
\csname url@samestyle\endcsname
\providecommand{\newblock}{\relax}
\providecommand{\bibinfo}[2]{#2}
\providecommand{\BIBentrySTDinterwordspacing}{\spaceskip=0pt\relax}
\providecommand{\BIBentryALTinterwordstretchfactor}{4}
\providecommand{\BIBentryALTinterwordspacing}{\spaceskip=\fontdimen2\font plus
\BIBentryALTinterwordstretchfactor\fontdimen3\font minus
  \fontdimen4\font\relax}
\providecommand{\BIBforeignlanguage}[2]{{%
\expandafter\ifx\csname l@#1\endcsname\relax
\typeout{** WARNING: IEEEtran.bst: No hyphenation pattern has been}%
\typeout{** loaded for the language `#1'. Using the pattern for}%
\typeout{** the default language instead.}%
\else
\language=\csname l@#1\endcsname
\fi
#2}}
\providecommand{\BIBdecl}{\relax}
\BIBdecl

\bibitem{Kantas2009}
N.~Kantas, ``{Sequential Decision Making in General State Space models},''
  Ph.D. dissertation, Cambridge, 2009.

\bibitem{Doucet2009}
A.~Doucet, N.~Kantas, S.~Singh, and J.~Maciejowski, ``{An Overview of
  Sequential Monte Carlo Methods for Parameter Estimation in General
  State-Space Models},'' in \emph{15th IFAC Symposium on System
  Identification}, no.~15.\hskip 1em plus 0.5em minus 0.4em\relax Saint Malo,
  France: Interntional Federation of Automatic Control, 2009.

\bibitem{Gordon1993}
N.~Gordon, D.~Salmond, and A.~Smith, ``{Novel approach to nonlinear and linear
  Bayesian state estimation},'' \emph{IEE Proceedings}, vol. 140, no.~2, pp.
  107--113, 1993.

\bibitem{West}
J.~Liu and M.~West, ``{Combined parameter and state estimation in
  simulation-based filtering},'' in \emph{Sequential Monte Carlo Methods in
  Practice}, A.~Doucet, N.~de~Freitas, and N.~Gordon, Eds.\hskip 1em plus 0.5em
  minus 0.4em\relax New York: Springer-Verlag, 2001, pp. 197--223.

\bibitem{Fearnhead2002}
P.~Fearnhead, ``{Markov chain Monte Carlo, Sufficient Statistics, and Particle
  Filters},'' \emph{Journal of Computational and Graphical Statistics},
  vol.~11, no.~4, pp. 848--862, Dec. 2002.

\bibitem{Storvik2002}
G.~Storvik, ``{Particle filters for state-space models with the presence of
  unknown static parameters},'' \emph{IEEE Transactions on Signal Processing},
  vol.~50, no.~2, pp. 281--289, 2002.

\bibitem{Carvalho2009}
C.~M. Carvalho, M.~Johannes, H.~Lopes, and N.~Polson, ``{Particle Learning and
  Smoothing},'' \emph{Statistical Science}, vol.~25, no.~1, pp. 88--106, 2010.

\bibitem{Whiteley2011a}
N.~Whiteley, A.~M. Johansen, and S.~Godsill, ``{Monte Carlo Filtering of
  Piecewise Deterministic Processes},'' \emph{Journal of Computational and
  Graphical Statistics}, vol.~20, no.~1, pp. 119--139, Jan. 2011.

\bibitem{doucetetal:2001:pffseojmls}
A.~Doucet, N.~Gordon, and V.~Krishnamurthy, ``Particle filters for state
  estimation of jump {Markov} linear systems,'' \emph{IEEE Trans. on Signal
  Proc.}, vol.~49, no.~3, March 2001.

\bibitem{blom}
H.~Blom and Y.~Bar-Shalom, ``The interacting multiple model algorithm for
  systems with markovian switching coefficients,'' \emph{IEEE Transactions on
  Automatic Control}, vol.~33, no.~8, pp. 780 --783, Aug 1988.

\bibitem{Fearnhead2007a}
P.~Fearnhead and Z.~Liu, ``{On-line Inference for Multiple Change Points
  Problems},'' \emph{Journal of the Royal Statistical Society, Series B},
  vol.~69, pp. 589--605, 2007.

\bibitem{yildirimandsinghanddoucet:2012:onlineem}
S.~Yildirim, S.~Sing, and A.~Doucet, ``An online {Expectation-Maximization}
  algorithm for changepoint models,'' \emph{Journal of Computational and
  Graphical Statistics, in press}, 2012.

\bibitem{nemethandfearnheadandmihaylovaandvorley:2012:fusion}
C.~Nemeth, P.~Fearnhead, L.~Mihaylova, and D.~Vorley, ``Bearings only tracking
  with joint parameter learning and state estimation,'' in \emph{Proceedings of
  the 15th International Conf. on Information Fusion}.\hskip 1em plus 0.5em
  minus 0.4em\relax ISIF, 2012, pp. 824--831.

\bibitem{nemethandfearnheadandmihaylovaandvorley:2012:ietconf}
------, ``Particle learning methods for state and parameter estimation,'' in
  \emph{Proc. of the 9th IET Data Fusion \& Target Tracking Conf.}\hskip 1em
  plus 0.5em minus 0.4em\relax 16-17 May, London, UK, 2012.

\bibitem{Arulampalam2002}
M.~S. Arulampalam, S.~Maskell, N.~Gordon, and T.~Clapp, ``{A tutorial on
  particle filters for online nonlinear/non-Gaussian Bayesian tracking},''
  \emph{IEEE Transactions on Signal Processing}, vol.~50, no.~2, pp. 174--188,
  2002.

\bibitem{Kalman1960}
R.~E. Kalman, ``{A New Approach to Linear Filtering and Prediction Problems},''
  \emph{Transactions of the ASME, Journal of Basic Engineering}, vol.~82, no.
  Series D, pp. 35--45, 1960.

\bibitem{Kong1994}
A.~Kong, J.~S. Liu, and W.~H. Wong, ``{Sequential Imputations and Bayesian
  Missing Data Problems},'' \emph{Journal of the American Statistical
  Association}, vol.~89, no. 425, pp. 278--288, Mar. 1994.

\bibitem{Carpenter1999}
J.~Carpenter, P.~Clifford, and P.~Fearnhead, ``{Improved particle filter for
  nonlinear problems},'' \emph{IEE Proceedings - Radar, Sonar and Navigation},
  vol. 146, no.~1, p.~2, 1999.

\bibitem{Douc2005}
R.~Douc and O.~Capp\'{e}, ``{Comparison of resampling schemes for particle
  filtering},'' in \emph{ISPA 2005. Proceedings of the 4th International
  Symposium on Image and Signal Processing and Analysis, 2005}, no.~4, 2005,
  pp. 64--69.

\bibitem{Kitagawa1996a}
G.~Kitagawa, ``{Monte Carlo Filter and Smoother for Non-Gaussian Nonlinear
  State Space Models},'' \emph{Journal of Computational and Graphical
  Statistics}, vol.~5, no.~1, pp. 1--25, Mar. 1996.

\bibitem{Pitt1999a}
M.~K. Pitt and N.~Shephard, ``{Filtering via Simulation: Auxiliary Particle
  Filters},'' \emph{Journal of the American Statistical Association}, vol.~94,
  no. 446, pp. 590--599, Jun. 1999.

\bibitem{Doucet2000}
A.~Doucet, S.~Godsill, and C.~Andrieu, ``{On sequential Monte Carlo sampling
  methods for Bayesian filtering},'' \emph{Statistics and computing}, vol.~10,
  no.~3, pp. 197--208, 2000.

\bibitem{Gilks2001}
W.~Gilks and C.~Berzuini, ``{Following a Moving Target-Monte Carlo Inference
  for Dynamic Bayesian Models},'' \emph{Journal of the Royal Statistical
  Society, Series B}, vol.~63, no.~1, pp. 127--146, 2001.

\bibitem{Bengtsson2008}
T.~Bengtsson, P.~Bickel, and B.~Li, ``{Curse-of-dimensionality revisited :
  Collapse of the particle filter in very large scale systems},''
  \emph{Probability and Statistics: Essays in Honor of David A. Freedman},
  vol.~2, pp. 316--334, 2008.

\bibitem{lindgren93}
B.~Lindgren, \emph{{Statistical Theory, Fourth Edition}}.\hskip 1em plus 0.5em
  minus 0.4em\relax Chapman \& Hall Texts in Statistical Science, 1993.

\bibitem{West1993}
M.~West, ``{Approximating posterior distributions by mixture},'' \emph{Journal
  of the Royal Statistical Society. Series B}, vol.~55, no.~2, pp. 409--422,
  1993.

\bibitem{Rong1993}
X.~{Rong Li} and Y.~Bar-Shalom, ``{Design of an Interacting Multiple Model
  Algorithm for Air Traffic Control Tracking},'' \emph{IEEE Transactions on
  Control Systems Technology}, vol.~1, no.~3, pp. 186--194, 1993.

\bibitem{julieranduhlmann:2004:ufane}
S.~Julier and J.~Uhlmann, ``Unscented filtering and nonlinear estimation,''
  \emph{Proceedings of the IEEE}, vol.~92, no.~3, pp. 401--422, 2004.

\bibitem{Russell2013}
Y.~Erol, L.~Li, B.~Ramsundar, and S.~J. Russell, ``The extended parameter
  filter,'' in \emph{Proceedings of the 30th International Conference on
  Machine Learning}, Atlanta, Georgia, 2009.

\end{thebibliography}
\bibliographystyle{apalike}

\end{document}